\documentclass[letter,apj]{emulateapj-rtx4}
\pdfoutput=1
\usepackage{graphicx}
\usepackage{enumerate}
\usepackage{amssymb, amsmath}
\usepackage{natbib}
\usepackage{color}
\usepackage{ulem}

\newcommand{\ias}{1}
\newcommand{\princeton}{2}
\newcommand{\sagan}{3}

\begin{document}

\title{Rotating Stellar Models Can Account for the Extended Main Sequence Turnoffs in Intermediate Age Clusters}
\author{Timothy D.~Brandt\altaffilmark{\ias, \sagan} \&
Chelsea X.~Huang\altaffilmark{\princeton}
}

\altaffiltext{\ias}{School of Natural Sciences, Institute for Advanced Study, Princeton, NJ, USA.}
\altaffiltext{\princeton}{Department of Astrophysical Sciences, Princeton University, Princeton, NJ, USA.}
\altaffiltext{\sagan}{NASA Sagan Fellow}

\begin{abstract}
We show that the extended main sequence turnoffs seen in intermediate age Large Magellanic Cloud (LMC) clusters, often attributed to age spreads of several hundred Myr, may be easily accounted for by variable stellar rotation in a coeval population.  We compute synthetic photometry for grids of rotating stellar evolution models and interpolate them to produce isochrones at a variety of rotation rates and orientations.  An extended main sequence turnoff naturally appears in color-magnitude diagrams at ages just under 1 Gyr, peaks in extent between $\sim$1 and 1.5 Gyr, and gradually disappears by around 2 Gyr in age.  We then fit our interpolated isochrones by eye to four LMC clusters with very extended main sequence turnoffs: NGC~1783, 1806, 1846, and 1987.  In each case, stellar populations with a single age and metallicity can comfortably account for the observed extent of the turnoff region.  The new stellar models predict almost no correlation of turnoff color with rotational $v \sin i$: the red edge of the turnoff is populated by a combination of slow rotators and edge-on rapid rotators.  
\end{abstract}

\section{Introduction} \label{sec:intro}

Exquisite photometry from the {\it Hubble Space Telescope} ({\it HST}) has recently revealed a surprise: many $\sim$1--2 Gyr-old star clusters in the Large Magellanic Cloud (LMC) show extended or multiple main sequence turnoffs (MSTOs).  First seen in NGC 1846 \citep{Mackey+Nielsen_2007}, and then also in NGC 1783 and 1806 \citep{Mackey+Nielsen+Ferguson+etal_2008, Goudfrooij+Puzia+Kozhurina-Platais+etal_2009}, these extended turnoff regions now seem to be a typical feature of intermediate-aged LMC clusters \citep[e.g.][]{Milone+Bedin_etal_2009}.  They cannot be explained away by photometric uncertainties or unresolved binaries, but can be reproduced by a spread of several hundred Myr in age \citep{Goudfrooij+Puzia+Kozhurina-Platais+etal_2011}.  

Large age dispersions are not expected in young and intermediate-aged globular clusters.  They have generally not been seen in younger massive LMC clusters \citep{Bastian+Silva_Villa_2013,Niederhofer+Hilker+Bastian_etal_2015}, though \cite{Correnti+Goudfrooij+Puzia+etal_2015} found NGC 1856 to be consistent with an 80 Myr age spread at $\sim$300 Myr.  Young globular clusters have long been modeled with coeval stellar populations of a single composition.  \cite{Elmegreen+Efremov_1997} showed that star formation in a typical $\sim$10$^5$ $M_\odot$ cluster can finish in $\lesssim$10$^7$ years, more than an order of magnitude smaller than the age spreads needed to explain the MSTOs in the LMC.  This picture of rapid star formation is confirmed by the hierarchical distribution of cluster ages seen in the LMC \citep{Efremov+Elmegreen_1998}.  Short bursts of star formation avoid the need to keep the cloud bound through the supernovae and copious ultraviolet radiation produced by the first generation of stars.  The existence of large age dispersions has also been challenged based on the narrowness of the subgiant branch and red clump \citep{Li+deGrijs+Deng_2014,Bastian+Niederhofer_2015}, though this conclusion is disputed \citep{Goudfrooij+Girardi+Rosenfield+etal_2015}.

Stronger evidence for multiple stellar populations has emerged for a few massive globular clusters.  The cluster $\omega$ Cen has two well-separated tracks in the color-magnitude diagram \citep{Bedin+Piotto+Anderson+etal_2004}, which can be interpreted as a very large helium enhancement in the smaller population \citep{Piotto+Villanova+Bedin+etal_2005}.  M2 also has several clearly separated tracks in a color-magnitude diagram \citep{Milone+Marino+Piotto+etal_2015}, each characterized by a different chemical abundance pattern \citep{Yong+Roederer+Grundahl+etal_2014}.  Similar results have been found for a handful of other Galactic globular clusters \citep[e.g.][]{Marino+Villanova+Piotto+etal_2008, Marino+Milone+Piotto+etal_2009}.

Many authors have explored rotation as an alternative explanation for the extended MSTO in intermediate-aged clusters.  Rotation has long been known to affect the MSTO due both to an increase in main sequence lifetime \citep{Meynet+Maeder_2000, Ekstrom+Georgy+Eggenberger+etal_2012} and to viewing angle differences \citep{vonZeipel_1924}.  Extreme rotators appear to be ubiquitous among early-type stars in the Solar neighborhood.  Vega ($\alpha$ Lyrae) is observed to rotate at $\sim$90\% of breakup \citep{Aufdenberg+Merand+Foresto+etal_2006, Peterson+Hummel+Pauls+etal_2006}; Altair, $\alpha$ Oph (Rasalhague), and $\alpha$ Cep (Alderamin), three of the brightest, closest A-type stars in the sky, are all similarly rapid rotators \citep{Monnier+Zhao+Pedretti+etal_2007, Zhao+Monnier+Pedretti+etal_2009}.  These stars are significantly deformed, with oblatenesses of $\sim$0.8--0.85 and large pole-equator differences in temperature.

The effect of orientation on the MSTO in intermediate-age LMC clusters was explored by \cite{Bastian+de_Mink_2009}, who found that it could explain at least some of the observed spread.  The ability of rotation to explain the entire spread was directly challenged by \cite{Girardi+Eggenberger+Miglio_2011}.  \cite{Girardi+Eggenberger+Miglio_2011} and \cite{Yang+Meng+Liu_2013} have both attempted to model LMC clusters with coeval populations of rotating stars.  \cite{Girardi+Eggenberger+Miglio_2011} found that two rotation rates could not explain the observed turnoffs, while \cite{Yang+Meng+Liu_2013} also found rotation to provide only a partial solution to the extended MSTO.  \cite{Goudfrooij+Girardi+Kozhurina-Platais+etal_2014} argued in favor of age rather than rotation on dynamical grounds, based on a correlation between turnoff width and central escape velocity (a proxy for the cluster's ability to retain gas).  \cite{Goudfrooij+Girardi+Kozhurina-Platais+etal_2014} also found a correlation between the prevalence of the bluest turnoff stars and of secondary red clumps, as expected from an age dispersion.

More recently, \cite{Brandt+Huang_2015b} noted that the nearby Hyades and Praesepe open clusters seem to be incompatible with a single nonrotating stellar population.  Stars at the top of the main sequence turnoff appear, at more than 99.99\% significance, to be younger than those at the base of the turnoff.  By interpolating the new rotating stellar models of \cite{Georgy+Ekstrom+Granada+etal_2013} as described in \cite{Brandt+Huang_2015}, \cite{Brandt+Huang_2015b} were able to fit both clusters with a single stellar population at a range of initial rotation rates and orientations.  We use similar techniques in this paper to investigate the effect of rotation on the MSTO in much richer LMC clusters.

We organize the paper as follows.  In Section \ref{sec:method}, we review our method of interpolating the rotating stellar models and performing synthetic photometry.  This procedure is described in much more detail in \cite{Brandt+Huang_2015}, hereafter BH15.  Section \ref{sec:prev_studies} compares our approach to previous attempts to explain the effect of rotation on the MSTO.  We describe our results in Section \ref{sec:results} and discuss their implications on the width of the subgiant branch and red clump, and on Galactic clusters, in Section \ref{sec:discussion}.  We conclude with Section \ref{sec:conclusions}, where we also suggest several areas for future research.

\section{Methodology} \label{sec:method}

Our methodology is described in detail in BH15.  Briefly, we interpolate rotating stellar models and perform synthetic photometry with a full gravity darkening calculation to obtain isochrones and color-magnitude diagrams.  We review the procedure here and refer the reader to BH15 for a more thorough discussion.  

We take the rotating stellar models of \cite{Georgy+Ekstrom+Granada+etal_2013} as our basic inputs.  We compute synthetic photometry using the gravity darkening model of \cite{Lara+Rieutord_2011}.  We integrate the specific intensities of the model atmospheres of \cite{Castelli+Kurucz_2004}, convolved with {\it HST} filter transmission curves \citep{Sirianni+Jee+Benitez+etal_2005}, over the projected stellar surface.  We then interpolate the rotating models in mass and metallicity using the fine nonrotating grid of \cite{Girardi+Bertelli+Bressan+etal_2002}.  We compute a rotational correction factor for each color and for each rotating model, and interpolate these factors linearly in magnitude (as power laws in flux) between the rotating models.  We then apply the interpolated rotational correction factors to nonrotating models at the same points in their evolution and scale the main sequence lifetimes to match those of the rotating models.  This procedure preserves the exact rotating models at the masses, metallicity, and rotation rates where they exist, and simply interpolates between them.

The \cite{Georgy+Ekstrom+Granada+etal_2013} rotating models are only available from 1.7 to 15 $M_\odot$; they resolve neither the transition from a convective to a radiative core, nor the transition from a radiative to convective envelope.  Both transitions happen below 1.7 $M_\odot$ and have a large impact on stellar rotation.  A convective core is uniform in composition, so that mixing at the outer core boundary efficiently supplies additional nuclear fuel to the central region where it is burned.  Stars with convective envelopes shed angular momentum in a magnetized wind and spin down rapidly.  Without the models to resolve this region of parameter space, we simply extrapolate the rotational correction to nonrotating models down to 1.45 $M_\odot$.  Our results at $\sim$2 Gyr ages, where the turnoff stars are $\sim$1.5 $M_\odot$, should therefore be viewed with caution.

BH15 constrained the age and composition of coeval star clusters by computing and multiplying the posterior probability distributions of individual stars.  This method neglects binaries, which could be observationally excluded from nearby clusters.  The clusters we study here are much too distant to resolve binaries either interferometrically or spectroscopically.  We do not attempt to generalize the method of BH15 to statistically account for unresolved binaries, deferring that to a future paper.

\section{Comparison to Previous Studies} \label{sec:prev_studies}

Several papers have previously attempted to explain the extended MSTOs seen in LMC clusters by a distribution of rotation rates and orientations.  \cite{Bastian+de_Mink_2009} suggested that the rapid rotation of massive stars could modify the morphology of isochrones and lead to an extended MSTO.  However, \citeauthor{Bastian+de_Mink_2009} only considered orientation effects and small differences in effective temperature, ignoring the substantial increase in main sequence lifetime that rotation can provide, and their results were directly challenged by \cite{Girardi+Eggenberger+Miglio_2011}.

\cite{Girardi+Eggenberger+Miglio_2011} performed stellar evolution calculations with two rotation rates, 0 and 150 km\,s$^{-1}$, and without core convective overshoot.  They also computed nonrotating models with a core overshooting parameter of 0.25.  They found that the combination of rotating and nonrotating tracks could not reproduce the extended MSTO of the LMC cluster NGC 1846.  Only a very wide distribution of ages combined with core overshooting provided a satisfactory fit.  
The results of \cite{Girardi+Eggenberger+Miglio_2011} contradict some of the approximations made by \cite{Bastian+de_Mink_2009}, who did not run full stellar evolution calculations.

More recently, \cite{Yang+Meng+Liu_2013} investigated the effects of rotation on the MSTO using their own stellar evolution tracks computed from the Yale Rotation Evolution Code \citep{Pinsonneault+Kawaler+Sofia+etal_1989}.  They found that stellar rotation could lead to extended MSTOs in clusters with ages between 0.8 and 2.2 Gyr, but could not account for the full width of the MSTO in NGC~1806.  \cite{Li+Mao+Chen_2015} and \cite{Li+Mao+Zhang_2015} attempted to reproduce the color-magnitude diagram of several clusters with a range of ages, binary fractions, and rotation rates; their best-fit models retained large spreads in age.  \cite{Li+deGrijs+Deng_2014} generated two evolutionary tracks from \cite{Georgy+Ekstrom+Granada+etal_2013} to explore the effect of rotation on post-main-sequence stars.  They concluded that the effect of rotation on the morphology and luminosities of the subgiant branch should be negligible due to the expansion of the stellar envelope.

In this work, we use the stellar evolution models of \cite{Georgy+Ekstrom+Granada+etal_2013}, taking the extended main sequence lifetimes and orientation effects fully into account.  Unlike the models used by \cite{Girardi+Eggenberger+Miglio_2011}, the \cite{Georgy+Ekstrom+Granada+etal_2013} tracks include modest core convective overshoot ($\alpha = 0.1$) in both the rotating and nonrotating cases, and cover a range of initial rotation rates.  We make no attempt to model the post-main-sequence portion of the color-magnitude diagram.  

Our approach is similar to that taken by \cite{Yang+Meng+Liu_2013}, though we adopt different stellar evolution calculations (see Section 2.3 of \citet{Yang+Meng+Liu_2013}).  \cite{Yang+Meng+Liu_2013} also only considered initial rotation periods of 0.49 days except in the case of NGC 1806, for which they assumed a Gaussian distribution centered on $\Omega_0/\Omega_{\rm crit} = 0.3$ with a dispersion of 0.06.  \cite{Li+Mao+Chen_2015} and \cite{Li+Mao+Zhang_2015}, like us, used the \cite{Georgy+Ekstrom+Granada+etal_2013} models.  However, they only applied a rotational correction to other stellar evolution models with a fixed distribution of rotation rates, and do not appear to have accounted for gravity darkening and orientation effects.

\section{Results} \label{sec:results}

\subsection{The Main Sequence Turnoff With Rotation}

While rotation has long been known to prolong stellar evolution \citep{Meynet+Maeder_2000}, large sets of stellar evolution models have only recently become available \citep{Ekstrom+Georgy+Eggenberger+etal_2012, Georgy+Ekstrom+Eggenberger+etal_2013, Georgy+Ekstrom+Granada+etal_2013}.  These models allow us to quantify the spread of the MSTO due to rotationally enhanced main sequence lifetimes.  

\begin{figure*}
\centering
\includegraphics[width=0.45\linewidth]{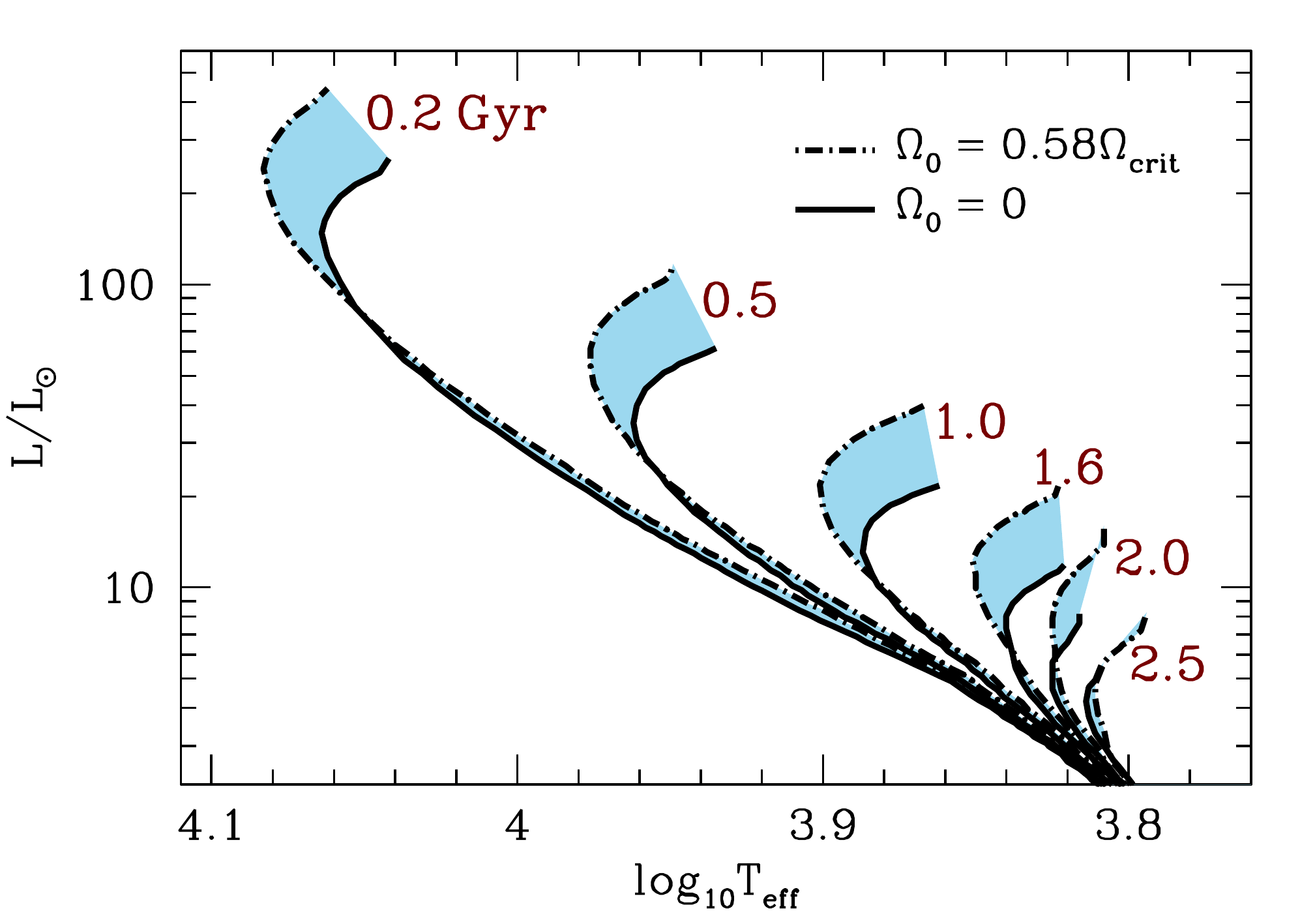}
\includegraphics[width=0.45\linewidth]{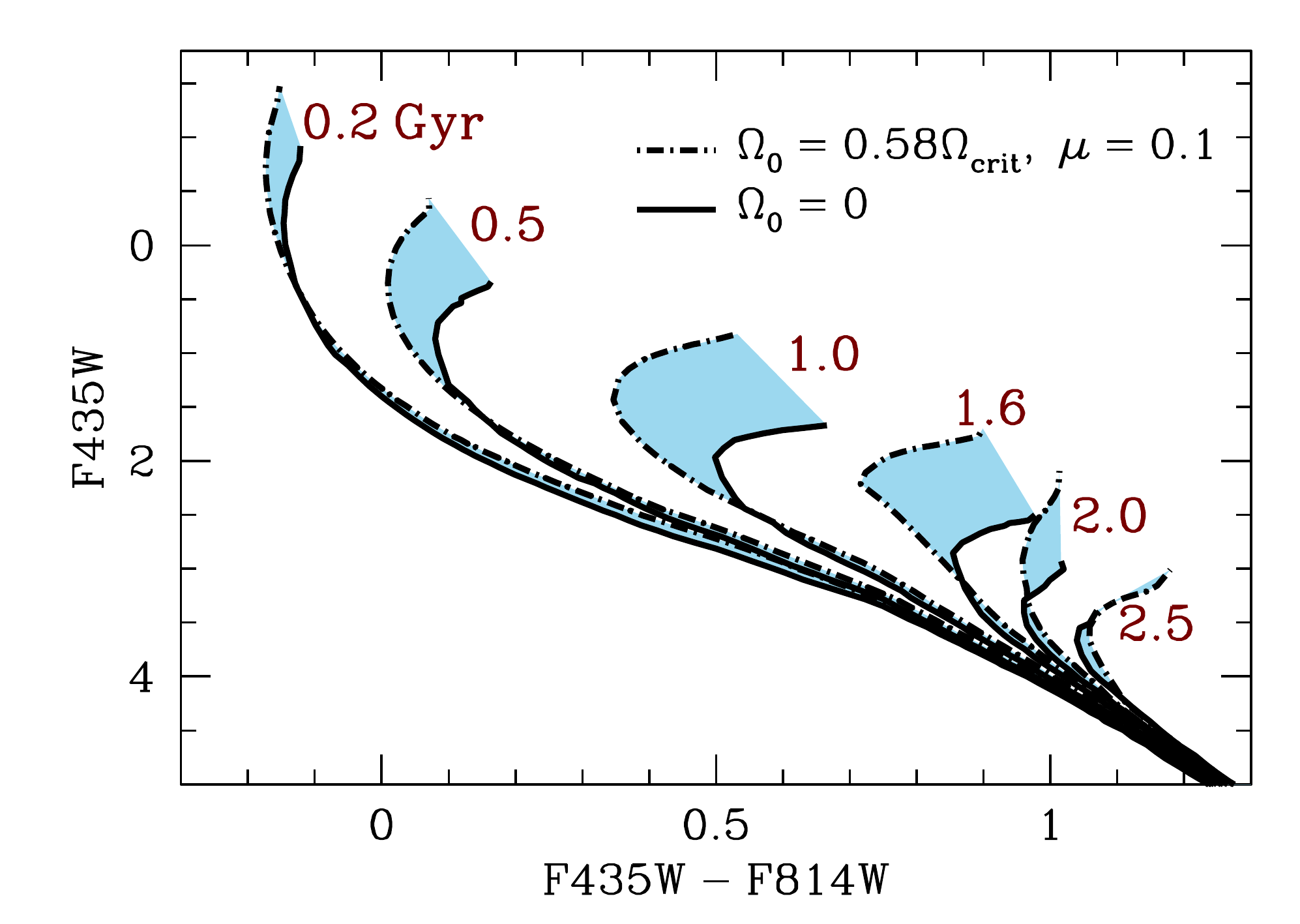}
\caption{The area bracketed by nonrotating and $\Omega_0/\Omega_{\rm crit} = 0.58$ isochrones at $Z_\odot$ \citep{Ekstrom+Georgy+Eggenberger+etal_2012}, in units of $\log L$ and $T_{\rm eff}$ (left panel) and in the {\it HST} filters $F435W$ and $F814W$ (roughly Johnson $B$ and $I$) at $\mu = \cos i = 0.1$ (right panel).  The shaded area in physical units is relatively constant with age until it disappears around 2 Gyr.  At this age, the turnoff stars have spent their main-sequence lives with convective outer envelopes, and those with rapid initial rotation have spun down.  The area in photometric units is not constant with age due to the nonlinear relationship between color and $T_{\rm eff}$.  The relation between color and $T_{\rm eff}$ also depends strongly on metallicity.  }
\label{fig:physical_HR}
\end{figure*}

Figure \ref{fig:physical_HR} shows the results of \cite{Ekstrom+Georgy+Eggenberger+etal_2012}, who computed tracks with initial rotation rates 0 and $\sim$58\% of critical, in physical units of $T_{\rm eff}$ and $\log L$ (left panel) and in synthetic {\it HST} photometry at $\cos i = 0.1$ ($i \approx 26^\circ$, right panel).  The main sequence turnoff has a pronounced extent in all populations younger than $\sim$2 Gyr.  At older ages, the turnoff stars had outer convective zones while on the main sequence and shed their angular momentum while young.  

The left panel of Figure \ref{fig:physical_HR} is not observable.  Colors are nonlinear functions of effective temperature and of metallicity; a turnoff region that is extended in physical units can be much less extended in photometry.  There is no reason for the extent of the main MSTO to remain roughly constant below $\sim$1.6 Gyr as it does in the left panel of Figure \ref{fig:physical_HR}.  Indeed, the right panel of Figure \ref{fig:physical_HR} shows a much less extended MSTO at ages $\lesssim$500 Myr.

\begin{figure}
\centering
\includegraphics[width=\linewidth]{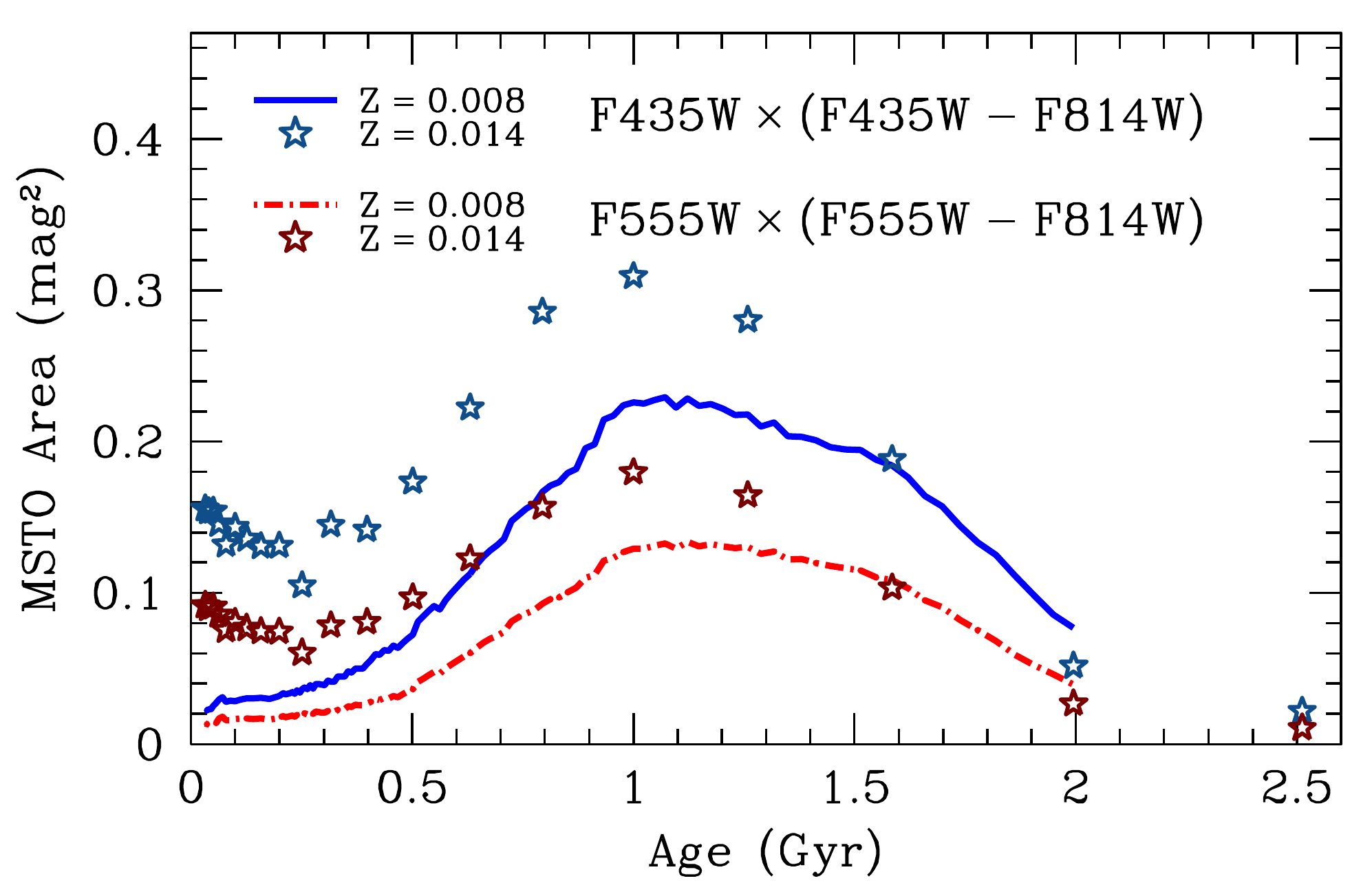}
\caption{Area bracketed by the nonrotating and $\Omega_0/\Omega_{\rm crit} = 0.6$, $\cos i = 0.9$ isochrones ($\Omega_0/\Omega_{\rm crit} = 0.58$ for $Z = Z_\odot = 0.014$), which we adopt as the extent of the MSTO.  
In both of the {\it HST} filter combinations shown, and particularly at low metallicity, the extent of the turnoff is very small below $\sim$500 Myr and peaks between 1 and 1.5 Gyr.  The details of the disappearance of the extended MSTO in the $Z=0.008$ isochrones at $\sim$1.5--2 Gyr should be viewed with caution.  The $Z_\odot$ isochrones \citep[open star symbols, from][]{Ekstrom+Georgy+Eggenberger+etal_2012} resolve the transition from convective cores and radiative envelopes to radiative cores and convective envelopes, but the low-metallicity isochrones (solid and dot-dashed lines) do not.
\label{fig:observed_HR}}
\end{figure}

Figure \ref{fig:observed_HR} shows the area of the MSTO on a color-magnitude diagram, with two choices of {\it HST} filter sets and with units of mag$^2$ (magnitude times color), at $Z_\odot$ and $Z=0.008$.  We define the turnoff area here as the area enclosed by a nonrotating isochrone and an isochrone rotating at $\Omega_0/\Omega_{\rm crit} = 0.6$ ($\Omega_0/\Omega_{\rm crit} = 0.58$ for $Z = Z_\odot = 0.014$) viewed at $\cos i = 0.9$ ($i = 26^\circ$).  An initial rotation rate 60\% of critical appears to be relatively common for $\sim$2 $M_\odot$ stars in the Solar neighborhood \citep{Zorec+Royer_2012}.  As noted in Section \ref{sec:intro}, Vega, Altair, $\alpha$ Cep, and $\alpha$ Oph all rotate much faster than this.  We show two {\it HST} filter combinations: $F435W$ and $F814W$ (solid blue line) and $F555W$ and $F814W$ (dot-dashed red line).  The three filters $F435W$, $F555W$, and $F814W$ correspond roughly to Johnson $B$, $V$, and $I$, respectively.

The MSTO in the physical Hertzsprung-Russell diagram (left panel of Figure \ref{fig:physical_HR}) is roughly constant in area on the log-log plot for ages $\lesssim$1.6 Gyr.  In the observable units of Figure \ref{fig:observed_HR}, the extent of the turnoff region has a clear peak between $\sim$1 and 1.5 Gyr, and is much less extended at both younger and older ages.  There is no extended turnoff at young ages because $B-I$ and $V-I$ depend weakly on $T_{\rm eff}$ at high temperatures, and especially at low metallicities.  The metallicity dependence means that the extended MSTO phenomenon should appear at a narrower range of ages in the LMC than in the more metal-rich Milky Way.  At old ages, the turnoff stars shed their initial angular momentum early in their main sequence lives and become effectively nonrotating.  

We caution that the decline shown in the turnoff area around 1.5--2 Gyr cannot be trusted in detail for the $Z=0.008$ isochrones.  The $Z_\odot$ models (open star symbols) clearly show this decline at $\sim$2 Gyr, but the low metallicity models only extend to 1.7 $M_\odot$.  We extrapolate them to 1.45 $M_\odot$ for our analysis, implicitly assuming that the effects of rotation change little from 1.7 to 1.45 $M_\odot$.  Depending on the details of the transition from a convective to a radiative core and from a radiative to convective envelope, the age at which the extended MSTO disappears could shift, and its disappearance could become more or less abrupt.  The $Z_\odot$ points, which {\it are} computed from full isochrones and are valid at all ages, show an even more dramatic falloff in MSTO area around 1.5--2 Gyr than the extrapolated $Z = 0.008$ models.

\subsection{Application to LMC Clusters}

We now attempt to reproduce observed intermediate age LMC clusters with very extended main sequence turnoffs, the best evidence for extended star formation in young globular clusters, with coeval stellar populations.  We simply interpolate the rotating isochrones of \cite{Georgy+Ekstrom+Granada+etal_2013} as described in Section \ref{sec:method} and perform our own synthetic photometry.  We adopt the rich clusters NGC 1783, NGC 1806, and NGC 1846 as our main comparison sample.  These were three of the first intermediate-aged clusters to show multiple MSTOs \citep{Mackey+Nielsen_2007, Mackey+Nielsen+Ferguson+etal_2008}, and show some of the most dramatically extended turnoffs.  We also show the younger, bluer cluster NGC 1987, which was claimed by \cite{Milone+Bedin_etal_2009} to have a higher metallicity, lower extinction, and larger fractional age spread than other LMC clusters.  

\begin{figure*}
\centering
\includegraphics[width=0.45\linewidth]{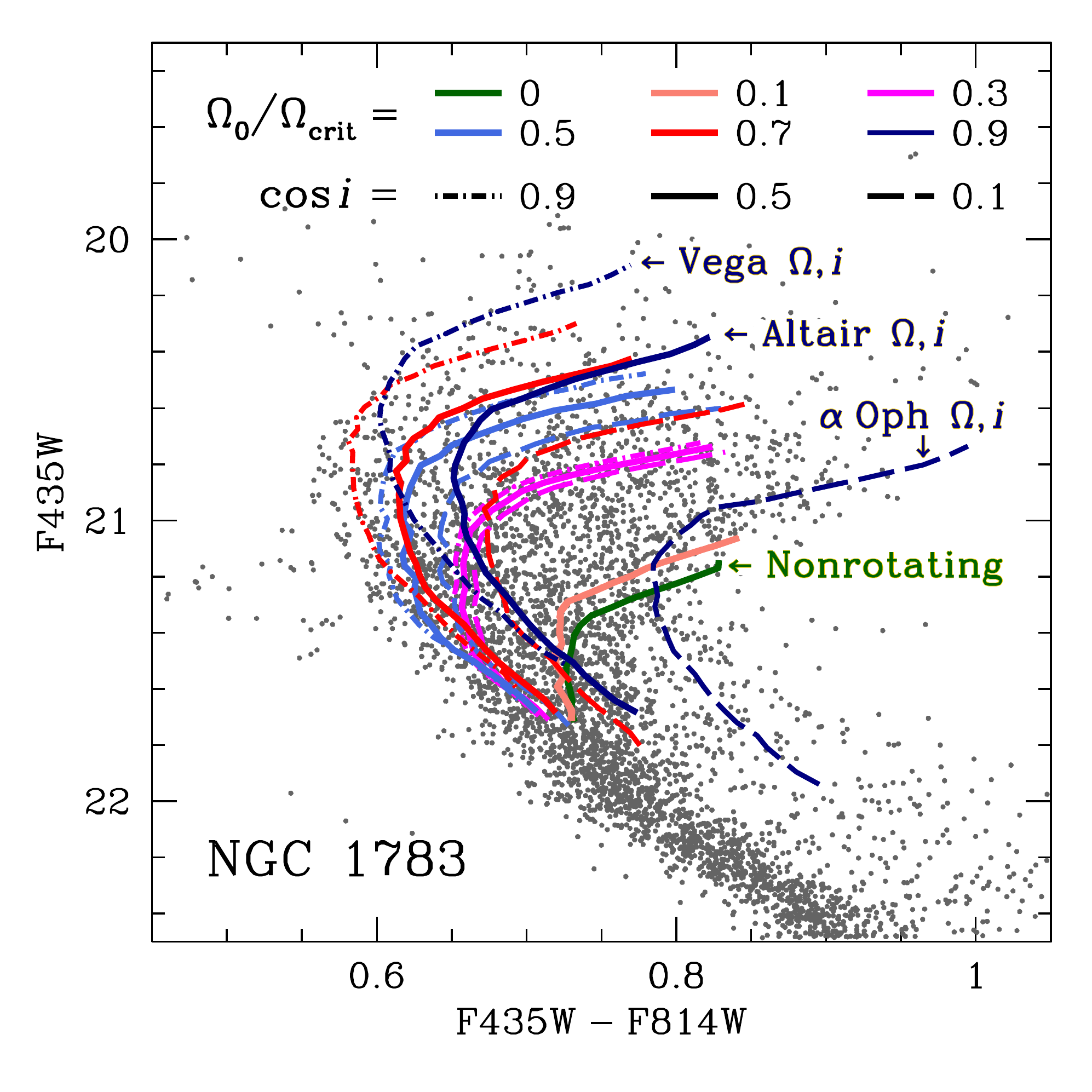}
\includegraphics[width=0.45\linewidth]{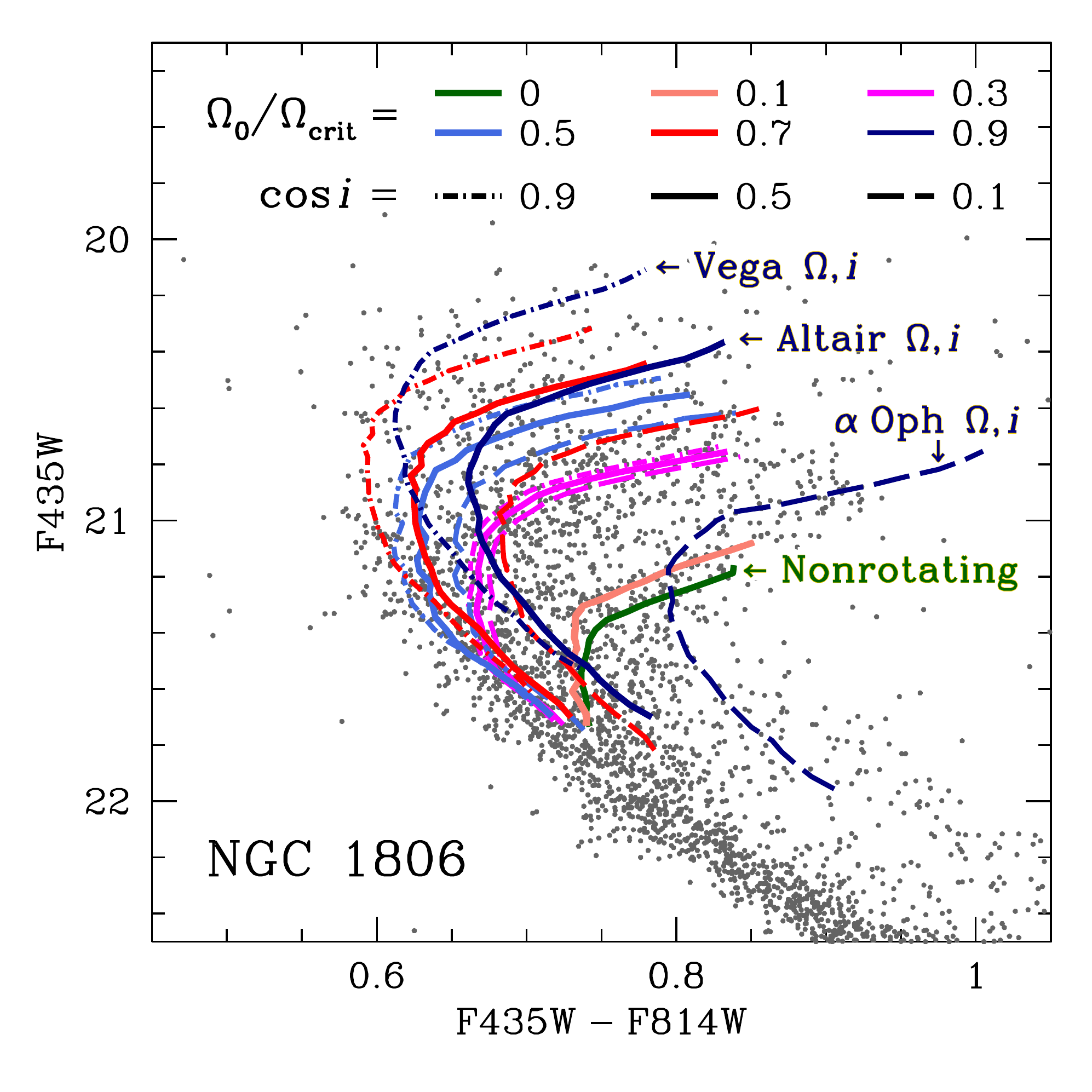}
\includegraphics[width=0.45\linewidth]{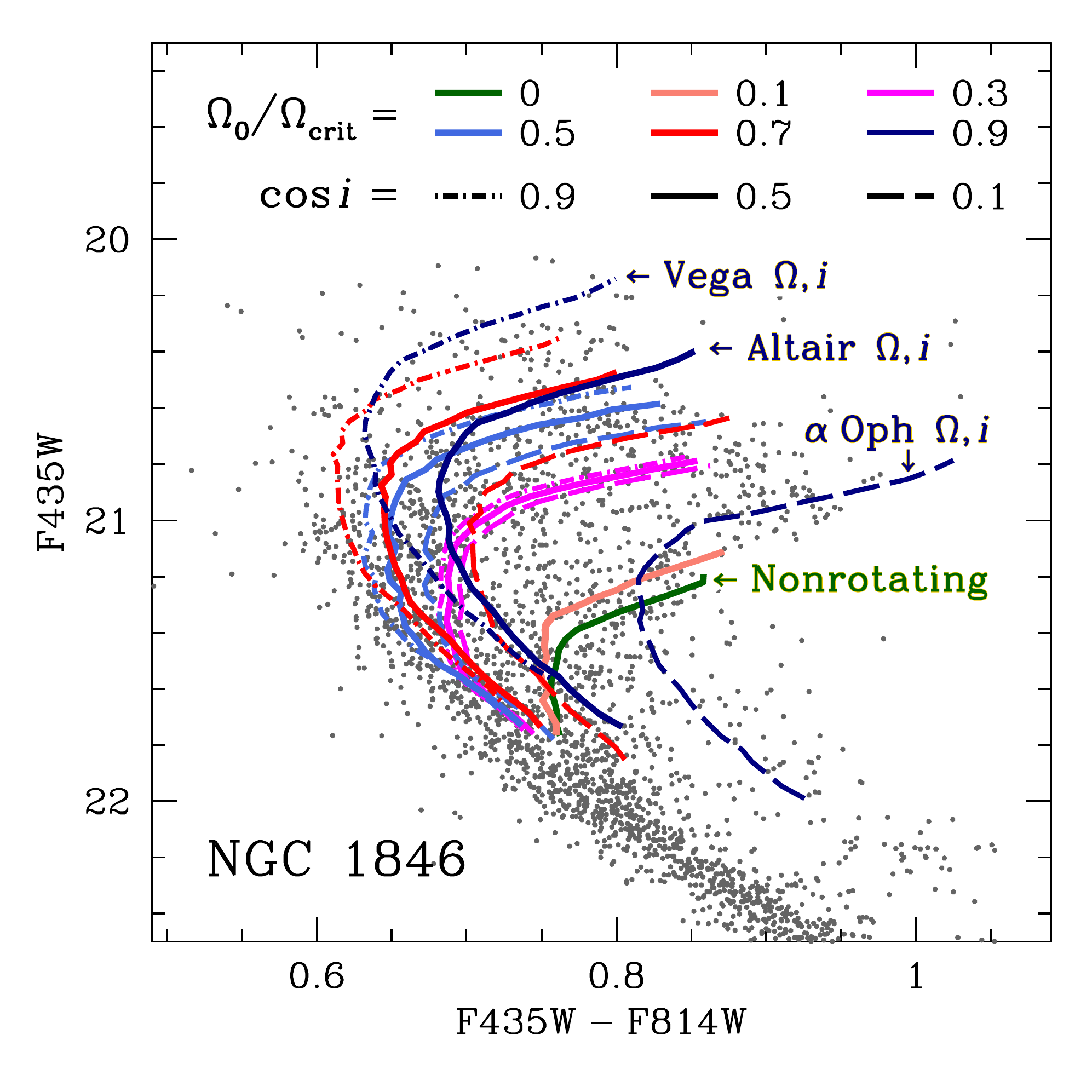}
\includegraphics[width=0.45\linewidth]{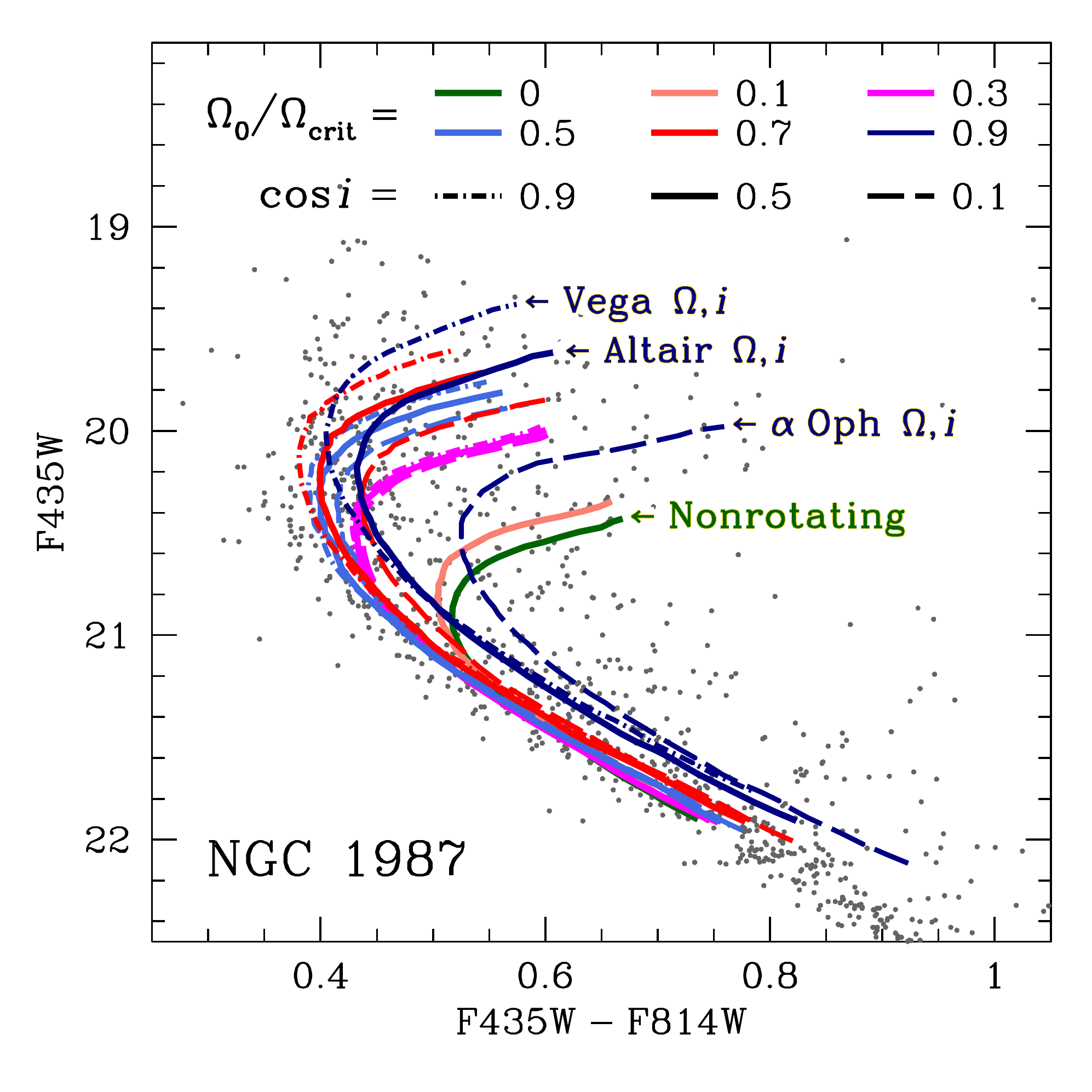}
\caption{Color-magnitude diagrams of four intermediate-age LMC clusters with strikingly extended MSTOs: NGC 1783 (top left), 1806 (top right), 1846 (bottom left), and NGC 1987 (bottom right).  The various curves are all coeval $Z=0.008$ isochrones at different initial rotation rates (colors) and viewing angles (line types).  Table \ref{tab:cluster_params} lists the assumed cluster ages, reddenings, and distances.  We have indicated bright, nearby A stars with $\Omega$ and $i$ corresponding to the most rapidly rotating, $\Omega_0/\Omega_{\rm crit} = 0.9$ isochrones; all three are among the 60 brightest stars in the sky, and the 12 brightest A stars.
A range of stellar rotation rates and viewing angles can comfortably account for an extended MSTO at a single age and metallicity.}
\label{fig:LMC_clusters}
\end{figure*}

Figure \ref{fig:LMC_clusters} overlays our interpolated isochrones on the decontaminated {\it HST} color-magnitude diagrams of NGC 1783, 1806, 1846, and 1987 as reduced and cleaned by \cite{Milone+Bedin_etal_2009}.
All isochrones have been shifted by hand using a reddening and distance modulus to bring them into rough agreement with the cluster observations.  We have made no attempt at a detailed parameter search.  The line colors represent initial rotation rates from zero to 90\% of critical, while the line types show viewing angles from $\cos i = 0.1$--0.9 ($i \approx 26^\circ$--84$^\circ$).  The three viewing angles are indistinguishable in the color-magnitude diagram for rotation rates $\Omega_0 \lesssim 0.3 \Omega_{\rm crit}$.

We emphasize that an initial rotation rate $\Omega_0/\Omega_{\rm crit} \approx 0.9$ does not appear to be exceptional among early-type stars near the Sun.  Vega, Altair, $\alpha$ Cep, and $\alpha$ Oph all rotate at about 90\% of breakup, with inclinations $\cos i \approx 0.01$, 0.5, 0.5, and 0.9, respectively \citep{Aufdenberg+Merand+Foresto+etal_2006, Monnier+Zhao+Pedretti+etal_2007, Zhao+Monnier+Pedretti+etal_2009}.  We have indicated the representative isochrones corresponding to these stars' values of $\Omega$ and $i$ on Figure \ref{fig:LMC_clusters}.

Table \ref{tab:cluster_params} summarizes our adopted cluster parameters.  All isochrones assume $Z=0.008$, reddening of $F435W-F814W = 0.09$ to 0.18, and true distance moduli of 18.60 to 18.65 mag assuming \cite{Fitzpatrick_1999} dust with $R_V = 3$.  This is larger by $\sim$0.1--0.15 mag than the best LMC distance \citep{Pietrzynski+Graczyk+Gieren+etal_2013}, a discrepancy of $\sim$5--7\% in linear distance.  Such a disagreement could be from, e.g., systematic errors in our synthetic photometry, an incorrect assumed metallicity, or incorrect assumed dust properties.  

\begin{deluxetable}{lcccr}
\tablewidth{0pt}
\tablecaption{Adopted Parameters for Four LMC Clusters}
\tablehead{
    NGC & 
    $(m - M)_0$ & 
    $E(B-I)$\tablenotemark{*} & 
    $Z$ & 
    Age (Gyr)}
\startdata
1783 & 18.65 & 0.09 & 0.008 & 1.55 \\
1806 & 18.65 & 0.10 & 0.008 & 1.55 \\
1846 & 18.65 & 0.12 & 0.008 & 1.55 \\
1987 & 18.60 & 0.18 & 0.008 & 1.00
\enddata
\tablenotetext{*}{More precisely, $E(F435W - F814W)$}
\label{tab:cluster_params}
\end{deluxetable}

Isochrones with a variety of rotation rates and viewing angles, but with a single age and composition, can easily reproduce the observed extent of the MSTO.  These results stand in stark contrast to those of \cite{Girardi+Eggenberger+Miglio_2011}, who concluded that rotation could not explain the turnoff of NGC 1846.  With a modest core overshoot parameter of 0.1 in all models \citep{Ekstrom+Georgy+Eggenberger+etal_2012, Georgy+Ekstrom+Granada+etal_2013}, rapidly rotating stars spend $\sim$25\% longer on the main sequence than nonrotating stars.  This accounts for most of the turnoff width and mimics a spread of $\sim$25\% in age (up to $\sim$400 Myr at the age of these clusters).  Orientation effects further increase the observed width of the turnoff region.  We note that a {\it fractional} age spread of $\sim$20--30\% is roughly consistent with the claims for LMC clusters \citep[80 Myr at an age of 300 Myr, and several hundred Myr at 1--1.5 Gyr][]{Milone+Bedin_etal_2009, Rubele+Girardi+Kozhurina-Platais+etal_2011, Rubele+Girardi+Kozhurina-Platais+etal_2013, Correnti+Goudfrooij+Puzia+etal_2015}.

\begin{figure*}
\includegraphics[width=0.5\linewidth]{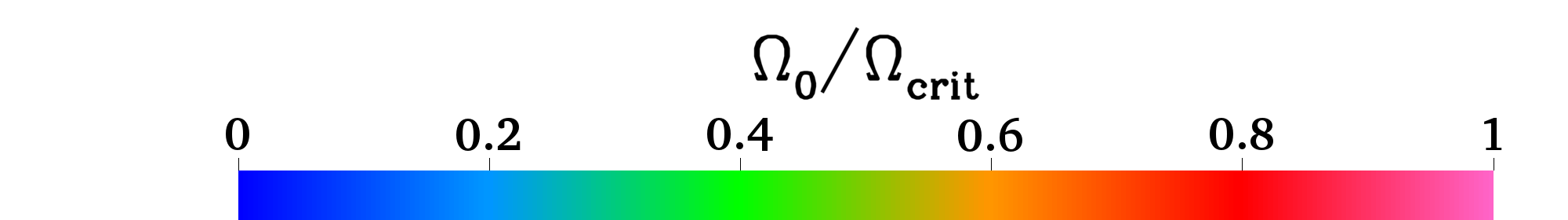}
\includegraphics[width=0.5\linewidth]{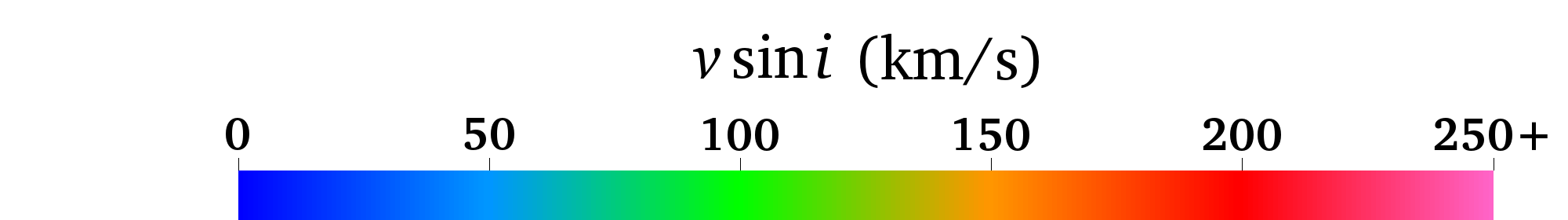}
\includegraphics[width=0.5\linewidth]{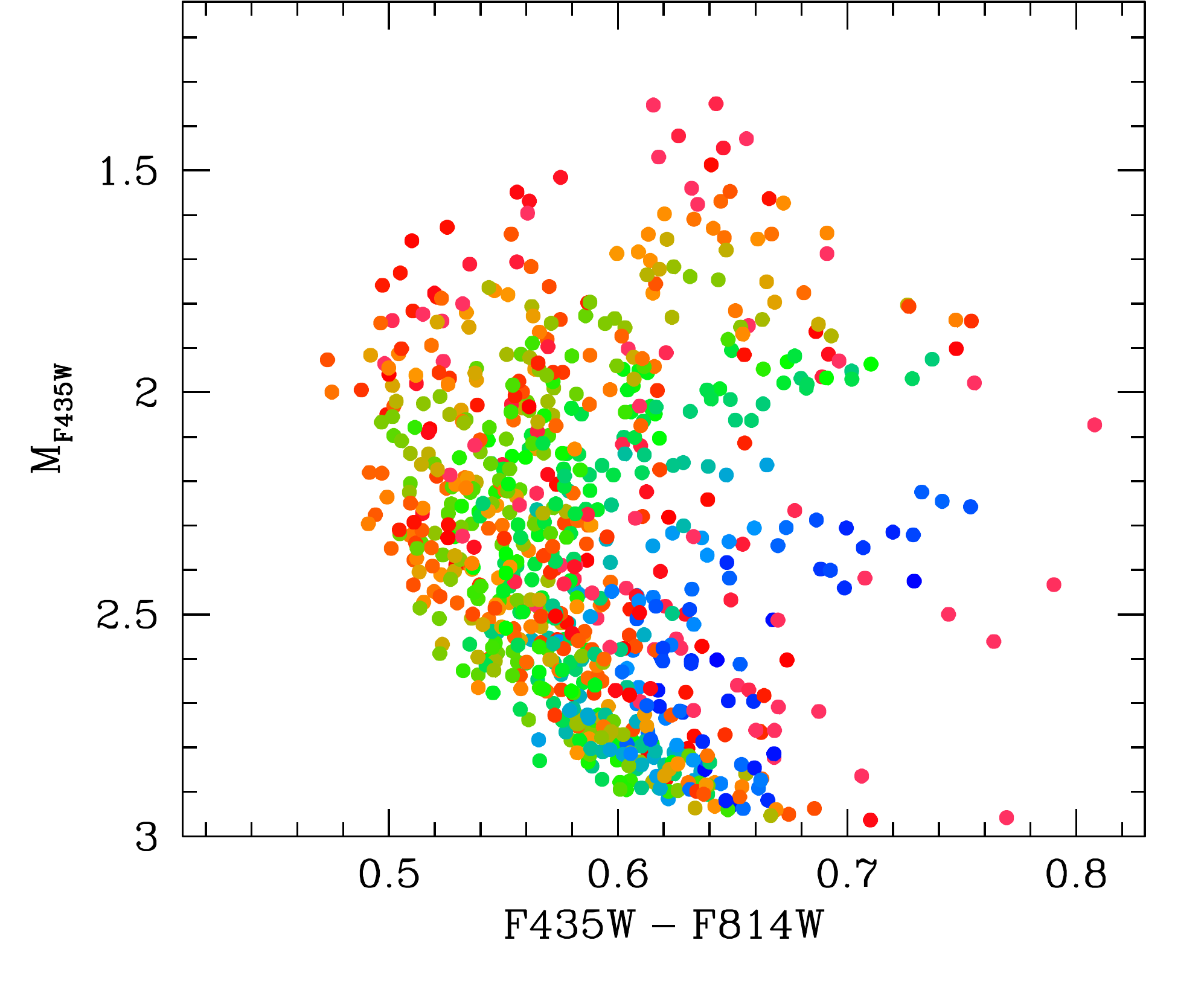}
\includegraphics[width=0.5\linewidth]{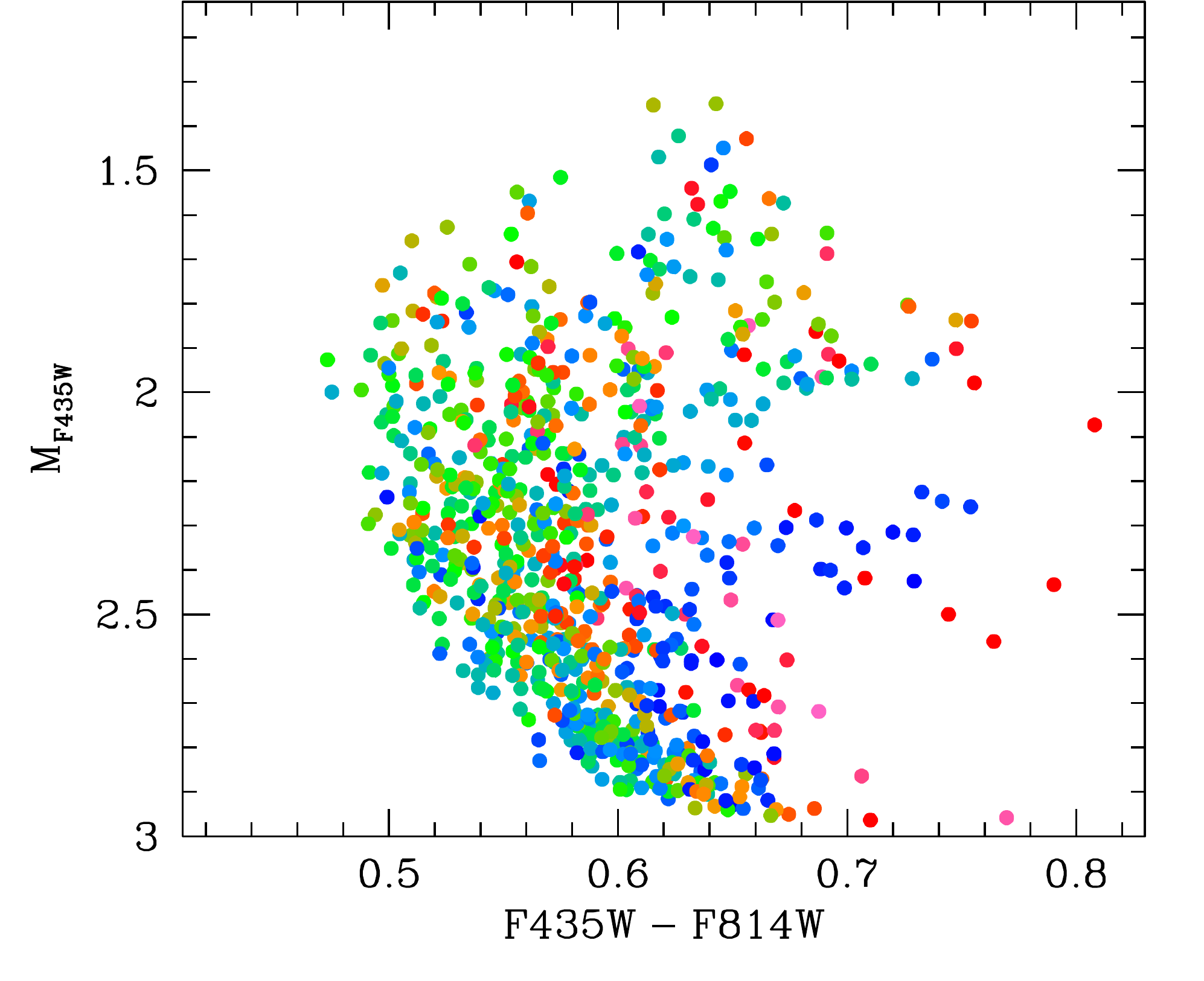}
\caption{Initial angular momenta $\Omega_0/\Omega_{\rm crit}$ (left panel) and current projected rotational velocities $v \sin i$ (right panel) for one realization of an unreddened, 1.55 Gyr-old, $Z=0.008$ stellar population with 10 mmag photometric errors in all bands.  We assume a Salpeter IMF and a Gaussian distribution of $\Omega_0/\Omega_{\rm crit}$ with a mean of 0.5 and a dispersion of 0.3 (truncated so that $\Omega_0/\Omega_{\rm crit} \in (0, 1)$) and neglect binaries.  The red side of the turnoff is populated by a combination of slow rotators and edge-on rapid rotators.  In contrast to \cite{Bastian+de_Mink_2009}, who identified red colors with rapid rotation, the models we present show no significant trend of $v \sin i$ with color.  The difference in mean $B-I$ color between stars with $v \sin i < 100$ km\,s$^{-1}$ and those with $v \sin i > 100$ km\,s$^{-1}$ is consistently $\lesssim$10 mmag for realizations of this cluster.  }
\label{fig:vsini}
\end{figure*}

Measurements of the projected rotational velocity $v \sin i$ could help to distinguish the effects of age and rotation.  \cite{Bastian+de_Mink_2009} identified the red end of the turnoff with the most rapid rotators, predicting a strong correlation of $v \sin i$ with color.  The models we present, however, show a much weaker correlation.  Figure \ref{fig:vsini} shows the initial angular momenta and current projected rotational velocities of one realization of a 1.55 Gyr-old, $Z=0.008$ population assuming 10 mmag (1\%) photometric errors.  We use a Gaussian distribution of $\Omega_0/\Omega_{\rm crit}$ centered on 0.5 with a dispersion of 0.3, in rough agreement with the values for nearby A and B stars \citep{Zorec+Royer_2012}.  We neglect the effect of gravity darkening on the inferred $v \sin i$ \citep{Fremat+Zorec+Hubert+etal_2005}.  

Unlike \cite{Bastian+de_Mink_2009}, we find the red end of the turnoff to be populated by a combination of slow rotators and edge-on rapid rotators, while the blue edge is comprised of pole-on rapid rotators.  The dispersion of $v \sin i$ increases with redder colors, but the correlation of color with mean $v \sin i$ is weak.  Intriguingly, this agrees with the weak correlation of $v \sin i$ with color seen in the $\sim$1.6 Gyr-old Galactic cluster Trumpler 20 \citep{Platais+Melo+Quinn+etal_2012}.  The difference in mean $B-I$ color between stars with $v \sin i < 100$ km\,s$^{-1}$ and those with $v \sin i > 100$ km\,s$^{-1}$ in Figure \ref{fig:vsini} is just 2 mmag.  Additional realizations of the same cluster show that these $B-I$ mean color differences are consistently $\lesssim$10 mmag.  As the left panel of Figure \ref{fig:vsini} shows, even the initial angular momentum $\Omega_0/\Omega_{\rm crit}$ is only moderately correlated with MSTO position.

\section{Discussion} \label{sec:discussion}

A major objection to age spreads in LMC clusters is the narrow distribution of stars along the post-main-sequence tracks \citep{Li+deGrijs+Deng_2014,Bastian+Niederhofer_2015}, though this result is disputed \citep{Goudfrooij+Girardi+Rosenfield+etal_2015}.  Rotating stellar models extend the MSTO region in two ways: by increasing the main sequence lifetime by a rotation-dependent factor, and by adding variation in color and apparent luminosity with viewing angle.  The former effect introduces a range of turnoff masses at fixed age, and thus a range of stellar masses on the subgiant branch and in the red clump.  Such a range of masses is exactly what would result from a spread in ages rather than rotation rates.  

Figure \ref{fig:LMC_clusters} shows that much of the extended MSTO phenomenon in our scenario arises because of the increase in lifetime with rotation (the line colors indicate different initial rotation rates).  Accounting for a broad MSTO with a spread in rotation rather than age can only reduce the spread in turnoff masses by a modest factor, corresponding roughly to the fraction of the spread attributable to orientation.  The difference in turnoff mass between a nonrotating and a very rapidly rotating isochrone is nearly 10\%, similar to the mass range produced by a 25\% age spread.

Another puzzle is the apparent lack of extended MSTOs in rich, $\sim$1.5 Gyr-old Galactic open clusters.  NGC 7789 has a slightly super-Solar metallicity \citep{Overbeek+Friel+Jacobson+etal_2015} and an estimated age of 1.4--1.6 Gyr \citep{Kalirai+Hansen+Kelson+etal_2008, Gim+Vandenberg+Stetson+etal_1998}, but does not have an obviously extended MSTO.  Trumpler 20, like NGC 7789, has a slightly super-Solar metallicity \citep{Carraro+Costa+Ahumada_2010} and an age $\sim$1.4--1.7 Gyr \citep{Carraro+Villanova+Monaco+etal_2014}.  Trumpler 20 does show hints of an extended MSTO corresponding to a $\sim$300 Myr spread in age \citep{Carraro+Villanova+Monaco+etal_2014}, though much of this may be due to differential reddening \citep{Platais+Melo+Quinn+etal_2012}.  

The lack of extended MSTOs in Galactic clusters could simply be due to the fact that the turnoff width is sharply declining at ages of 1.5--1.8 Gyr at $Z_\odot$ (Figure \ref{fig:observed_HR}).  NGC 7789 and Trumpler 20 have super-Solar metallicities, and could plausibly show a decline in turnoff width at younger ages (equivalently, at lower turnoff masses).  The rotating models we use also suggest that these clusters are slightly older than has been reported in the literature; the expected MSTOs would then be correspondingly narrower.  

One way to simultaneously account for a narrow subgiant branch and for different turnoff behaviors in the Galaxy and the LMC is with systematically faster rotation at low metallicity.  The opacity in stars and in molecular clouds is dominated by heavy elements.  This opacity, in turn, sets the temperature gradient at which convection begins.  Changing the heavy element abundance by a factor of three, roughly the difference between young Galactic open clusters and the LMC, could plausibly have a strong effect on the dissipation of angular momentum.  A bias toward rapid rotation at low $Z$ would increase the relative importance of orientation in accounting for the turnoff width.  It would thus allow for a narrower distribution of turnoff masses, while also narrowing the MSTO in higher-metallicity populations.

\section{Conclusions and Future Work} \label{sec:conclusions}

Our results show that variable stellar rotation and orientation at fixed age and composition can easily account for extended MSTOs, even in those LMC clusters with the most dramatically extended turnoffs.  Stellar rotation naturally explains the appearance of a broad MSTO at ages just younger than 1 Gyr and its disappearance at ages of $\sim$2 Gyr.  \cite{Brandt+Huang_2015b} showed that rotation can also remove a spread in color-magnitude ages in the much closer and less massive Hyades and Praesepe clusters.  Taken together, these results strongly support the formation of open clusters in short bursts of star formation, forming essentially coeval stellar populations of uniform composition.  We suggest several areas of further study to validate and refine this picture.  

In principle, with a sufficiently fine grid of stellar models and a careful accounting of stellar binaries, we can invert the observed distribution of stars in the color-magnitude diagram to obtain the distribution of initial rotation rates.  We have made no attempt to do so in this paper.  Such an inversion would require marginalizing over parameters including reddening, distance, binarity, metallicity, age, and rotation, a highly complex process with strong covariances between the parameters.  The $\sim$0.1--0.15 mag difference between the best LMC distance and the distance moduli used in Figure \ref{fig:LMC_clusters}, though modest, suggests that we would need to treat any result with caution.  

Another promising avenue for future work is a detailed study of the disappearance of the extended MSTO phenomenon at ages $\sim$2 Gyr.  We are unable to resolve this transition in the low-metallicity case because the \cite{Georgy+Ekstrom+Granada+etal_2013} rotating stellar models only extend to 1.7 $M_\odot$.  We extrapolate them down to 1.45 $M_\odot$, but in order to resolve the change in MSTO morphology, we would need models at many rotation rates extending at least to masses of $\sim$1.3 $M_\odot$.  Two important transitions, from a convective to radiative core and from a radiative to convective envelope, occur below 1.7 $M_\odot$.  Both transitions change the effect of rotation on stellar evolution.  Our extrapolation, even down to 1.45 $M_\odot$, makes it difficult to have much confidence in the form of the decline in the turnoff area around 2 Gyr in Figure \ref{fig:observed_HR}.  The full isochrones at $Z_\odot$ and at two rotation rates, the open stars in Figure \ref{fig:observed_HR}, show that the disappearance of an extended MSTO at $\sim$2 Gyr may be even more abrupt than our $Z=0.008$ extrapolation indicates.

Finally, we suggest that the resolution of the disagreement between the range of ages suggested by the extended MSTOs and by the observed spread in the subgiant branch will require a detailed study.  A range of rotation rates at constant age and composition will produce a spread in masses at a fixed point in the turnoff region, just like a spread in ages.  The mass ranges in these two scenarios are likely to be comparable and as large as $\sim$10\%.  The extent to which rotation can reduce the spread in turnoff masses depends on the detailed behavior of the stellar models with rotation and on the distribution of initial rotation rates.  Both deserve a much more detailed analysis than that presented here.

\acknowledgements{The authors thank an anonymous referee for helpful suggestions, particularly on the expected distribution of projected rotational velocities.  This work was performed in part under contract with the Jet Propulsion Laboratory (JPL) funded by NASA through the Sagan Fellowship Program executed by the NASA Exoplanet Science Institute. }

\bibliographystyle{apj_eprint}
\bibliography{refs}

\end{document}